\documentclass{article}
\usepackage{spconf,amsmath,graphicx}

\usepackage{enumitem}
\setlist{nosep, leftmargin=14pt}

\usepackage[algoruled, linesnumbered]{algorithm2e}

\usepackage{xcolor}

\definecolor{linkred}{HTML}{a11616}
\definecolor{linkgreen}{HTML}{16a116}
\definecolor{linkblue}{HTML}{1616a1}
\usepackage[draft]{hyperref}
\hypersetup{
    colorlinks=true,
    linkcolor=linkred,
    filecolor=linkblue,
    urlcolor=linkblue,
    citecolor=linkgreen
}

\usepackage{booktabs}
\usepackage{multirow}

\newcommand{\norm}[1]{\left\|#1\right\|}
\newcommand{\SVT}{\operatorname{SVT}}
\newcommand{\soft}{\operatorname{soft}}

\newcommand{\data}{\mathbf{d}}  
\newcommand{\adjoint}[1]{#1^*}
\renewcommand{\L}{\mathbf{L}}
\renewcommand{\S}{\mathbf{S}}
\newcommand{\X}{\mathbf{X}}
\newcommand{\M}{\mathbf{M}}
\newcommand{\N}{\mathbf{N}}

\newcommand{\lambdaL}{\lambda_{\text{L}}}
\newcommand{\lambdaS}{\lambda_{\text{S}}}

\newcommand{\E}[2]{#1\!\cdot\!10^{#2}}

\usepackage{tikz, pgfplots}
\pgfplotsset{compat=newest}
\usetikzlibrary{arrows.meta,bending,positioning}
\graphicspath{{figures/}}

\usepackage{caption}
\usepackage{subcaption}

\newcommand{\Ktrans}{$K_\textup{trans}$}
\newcommand{\Fp}{$F_\textup{p}$}
\newcommand{\Tc}{$T_\textup{c}$}

\title{Improving DCE-MRI through unfolded low-rank + sparse optimisation}
\name{Ondřej Mokrý$^{1}$ \qquad Jiří Vitouš$^{2, 3}$ \qquad Pavel Rajmic$^{1}$ \qquad Radovan Jiřík$^{3}$}

\address{%
    $^{1}$ Brno University of Technology, 
               Dept.\ of Telecommunications, 
               Technická 12,
               Brno,
               Czech Republic \\
    $^{2}$ Brno University of Technology, 
               Dept.\ of Biomed.\ Engineering, 
               Technická 12,
               Brno,
               Czech Republic \\
    $^{3}$ Institute of Scientific Instruments of the CAS,
                 Královopolská 147,
                 Brno, Czech Republic}
\begin{document}
%
\maketitle
\begin{abstract}
A method for perfusion imaging with DCE-MRI is developed based on two popular paradigms:
the low-rank + sparse model for optimisation-based reconstruction,
and the deep unfolding.
A~learnable algorithm derived from a proximal algorithm is designed with emphasis on simplicity and interpretability.
The resulting deep network is trained and evaluated using a simulated measurement of a rat with a brain tumor, showing large performance gain
over the classical low-rank + sparse baseline.
Moreover, quantitative perfusion analysis is performed based on the reconstructed sequence,
proving that even training based on a simple pixel-wise error can lead to significant improvement of the quality of the perfusion maps.
\end{abstract}
%
%
%
%
\section{Introduction}
\label{sec:intro}


Dynamic Contrast-Enhanced (DCE) Magnetic Resonance Imaging (MRI) is an experimental technique for measuring perfusion parameters of tissues, such as blood plasma flow, blood plasma volume and vessel-wall permeability. These parameters characterize the microvascular system and are known as important biomarkers e.g.\ for staging of tumors and monitoring of tumor therapy, as tumor's microvascular function and structure react on a~treatment significantly faster than the commonly measured tumor volume.
DCE-MRI is based on intravenous administration of a contrast agent and rapid imaging of the tissue of interest, capturing the spatial and temporal distribution of the contrast agent.
The acquired image sequence is then
used to fit
a pharmacokinetic model parametrized by the sought perfusion parameters. Use of advanced pharmacokinetic models requires high temporal resolution (approx.\ one frame per second) which is beyond the limits of standard MRI. This motivates advance in accelerated DCE-MRI, recently mostly based on compressed sensing and deep learning (DL).

In compressed-sensing DCE-MRI, the sparsely sampled data are reconstructed by employing prior information in the form of spatio-temporal regularisation,
formulated using for instance the total variation \cite{Chambolle2004:TV.algorithm, Schloegl_ICTGV}, low-rank \cite{Dankova2016, MangovaTSP2017} or their combination in low-rank plus sparse (L+S) model \cite{Candes:2011Robust.princ.comp.anal,OtazoCandesSodickson:MRIperfusion,DankovaRajmicJirik2015:LVA}. 
Another possibility to reconstruct sparsely sampled data is based on end-to-end DL methods \cite{Schlemper2017:Deep.cascade.of.CNN.for.MRI}.
Such methods do not employ any information about the physical process of MRI acquisition, tend to 
overfitting and to the so called hallucinations (imposing of image structures from the training dataset but not present in the currently imaged object) \cite{Knoll2020:Deep.Learning.Parallel.MRI.Survey,Liang2020:Deep.MRI}. 
This is why physics-informed DL approaches are preferred by the MRI community, often formulated in the form of unfolded (unrolled) algorithms \cite{Hammernik2017:Learning.Variational.Network.MRI,Huang2021:Deep.L.plus.S.network}.
These methods are, however, still quite general, require large training datasets and still suffer from possible overfitting.

Here we suggest a method that keeps the role of the L~and S components from the L+S approach, possibly generalises the regularisation used in the baseline method, and allows learning of the (generalised) regularisation parameters.
Thus by connecting the L+S with DL, we aim at a~straightforward interpretation and enable learning on only limited-size datasets (which is a common restriction in DCE-MRI).

Preliminary work concerning the unfolded L+S model for DCE-MRI was published as \cite{Mokry2022:MRI.EEICT}.
Even though the present work is based on the same L+S model, it considers several parametrisation options (instead of just the simple one mentioned in Sec.\,\ref{ssec:parametrisation} below),
both the simulation and learning protocols were optimised;
importantly, the perfusion analysis is newly included in the evaluation.


\section{L+S model unfolded}
\label{sec:LS}

\subsection{The model and its solution}

Low-rank plus sparse model is a~popular approach to data modelling.
The model is based on the assumption that the  acquired data can be split into two differently behaving components.
In the case of DCE-MRI, the sequence of images can be split into a~constant or slowly varying component (background) and a~component with faster dynamics.
Mathematically, these components are characterised by their low-rankness and sparsity, respectively.

The usual optimisation problem used for L+S modelling reads
\cite{OtazoCandesSodickson:MRIperfusion}
\begin{equation}
    \label{eq:LS_reco_model}
    \arg\min_{\L, \S} \ \tfrac{1}{2}\norm{\mathcal{A}(\L+\S)-\data}_2^2 + \lambdaL\norm{\L}_* + \lambdaS\norm{\mathcal{T}\S}_1\!.
\end{equation}
Here, $\L$ and $\S$ are the (sought) low-rank and sparse components of the image sequence $\L+\S$,
acquired by the MR scanner as $\data$ using the linear imaging operator
$\mathcal{A}$, which includes the coil-sensitivity-aware encoding of the data into k-space and subsampling.
The linear operator $\mathcal{T}$ transforms
$\S$
into a~space where it is sparse;
typically, $\mathcal{T}$ represents pixel-wise differences in the time direction.
The norms used are the Euclidean norm for the data term, indicating that Gaussian noise is assumed in the acquisition process;
$\norm{\L}_*$ is the nuclear norm forcing $\L$ to be low-rank
\cite{Recht2010:nuclear.norm};
$\norm{\mathcal{T}\S}_1$ is the $\ell_1$ norm forcing $\mathcal{T}\S$ to have as many zero components as possible
\cite{DonohoElad2003:Optimally}.
Finally, the
constants $\lambdaL,\lambdaS>0$
have to be chosen by the user, and balance the weight of the three terms in \eqref{eq:LS_reco_model}.

The problem \eqref{eq:LS_reco_model} can be solved numerically using proximal splitting algorithms.
In our case, we exploit the primal-dual algorithm of Chambolle and Pock (CPA)
\cite{ChambollePock2011:First-Order.Primal-Dual.Algorithm},
see Alg.~\ref{alg:CP}.
%
%
The operator $\soft_{\alpha}$ denotes the usual soft thresholding with threshold $\alpha>0$.
%
%
%
%
The singular value thresholding (SVT) \cite{Cai2010:SVT} with threshold $\alpha$, denoted $\SVT_\alpha$, is defined via the singular value decomposition $\X = \mathbf{U}\operatorname{diag}(\boldsymbol\sigma)\mathbf{V}^{\textup{H}}$ as $\SVT_\alpha(\X) = \mathbf{U}\operatorname{diag}\left(\soft_\alpha(\boldsymbol\sigma)\right)\mathbf{V}^{\textup{H}}$.
%
%
The algorithm can be stopped after a~finite number of iterations or it can terminate after a~heuristic convergence criterion is met.

\vspace{-6pt}
\begin{algorithm}
    \small
    \caption{%
    CPA
    solving the problem \eqref{eq:LS_reco_model}}
    \label{alg:CP}
    choose $\rho,\tau >0$ such that $\rho\tau < 1/(4\norm{\mathcal{A}}^2 + \norm{\mathcal{T}}^2)$ \\
    set initial iterates $\bar\L^{(0)}$, $\bar\S^{(0)}$, $\M^{(0)}$, $\N^{(0)}$ \\
    \For{$k=0,1,\dots$}
    {
        $\M^{(k+1)} =
        (\M^{(k)} + \rho\mathcal{A}(\bar\L^{(k)}+\bar\S^{(k)}) - \rho\data)/(1 + \rho)\hspace{-2em}$ \\
        $\N^{(k+1)} =
        \N^{(k)} + \rho \mathcal{T}\bar\S^{(k)} - \soft_{\lambdaS}(\N^{(k)} + \rho \mathcal{T}\bar\S^{(k)})$%
        \label{ln:L.update} \\
        $\L^{(k+1)} =
        \SVT_{\tau\lambdaL}(\L^{(k)} - \tau\adjoint{\mathcal{A}}\M^{(k+1)})$%
        \label{ln:S.update} \\
        $\S^{(k+1)} =
        \S^{(k)} - \tau\adjoint{\mathcal{A}}\M^{(k+1)} - \tau \mathcal{T}^\top\N^{(k+1)}$ \\
        $\bar\L^{(k+1)} = 2\L^{(k+1)}-\L^{(k)}$ \\
        $\bar\S^{(k+1)}  = 2\S^{(k+1)} - \S^{(k)}$
    }
    \Return{$\L^{(k+1)}$, $\S^{(k+1)}$}
\end{algorithm}
\vspace{-14pt}

\subsection{Deep unfolding and parametrisation}
\label{ssec:parametrisation}

We choose the paradigm of deep unfolding \cite{Liu2019:Deep.Proximal.Unrolling,MongaLiEldar2021:Algorithm.unrolling} to form a~model-based network for solving the L+S model.
In such an approach, a~fixed number $K$ of iterations of an iterative algorithm is considered to represent $K$ layers of a~network.
To enable learning, a~part of each iteration/layer is then replaced by a~learnable component,
supposedly the same one in each iteration to preserve similarity with the iterative template.
An even closer similarity to the original algorithm may be achieved by forcing the related learned parameters to be equal across the layers of the network; we refer to this as the \emph{tied} variant of a~network.
With a~proper choice of the learnable component, it may be possible to show its correspondence to a~proximal operator \cite{Gribonval2020:Characterization.of.prox},
or even to derive an optimisation problem which is solved by such a network.
In contrast, the \emph{untied} variant of the network provides more degrees of freedom,
and is therefore expected to deliver better results at the cost of increased computational complexity.

A crucial part of the design of the unfolded network is the choice of parameters to be optimised.
Since we aim at preserving the interpretation of the two components of the solution,
we choose simple (yet effective) parametrisations of the soft thresholding (or similar) operator (line \ref{ln:L.update} and \ref{ln:S.update} of Alg.~\ref{alg:CP}):
\begin{itemize}
    \item The most basic option is to only use learning to find
    the optimal $\lambdaL,\lambdaS$ for problem \eqref{eq:LS_reco_model}.
    These parameters then appear as thresholds in the soft-thresholding and SVT,
    as is visible in Alg.~\ref{alg:CP}.
    We refer to this approach as the \emph{simple} activation.
    \item The \emph{soft} activation allows to adapt not only the threshold, but also the slope of the function above this threshold (the slope is fixed to $1$ in the \emph{simple} case).
    \item Non-negative garrote \cite{Breiman1995:Nonnegative.garrote} is an operator similar to soft thresholding
    zeroing-out the values below a given threshold (which is learnable in our case), but for large input values, it tends towards the identity (in contrast to soft thresholding which has a constant bias for inputs exceeding the threshold).
\end{itemize}
Note that all the above cases use a~very low number of parameters, compared to usual neural networks.
More specifically, the number of parameters can only get up to $4K$ in the \emph{soft} case for a $K$-layer model.
However, this is in line with the concept of adjusting the L+S model using little training data, rather than training a~dedicated complex architecture.

\section{Experiments and results}
\label{sec:exp}

\subsection{Data simulation}
\label{ssec:simulation}

\begin{figure}
    \centering
    \scalebox{0.50}{
%
%
\definecolor{mycolor1}{rgb}{0.85098,0.32549,0.09804}%
\definecolor{mycolor2}{rgb}{0.92941,0.69412,0.12549}%
\definecolor{mycolor3}{rgb}{0.46667,0.67451,0.18824}%
\definecolor{mycolor4}{rgb}{0.00000,0.44706,0.74118}%
\begin{tikzpicture}

\begin{axis}[%
width=3.702in,
height=3.702in,
at={(0in,0in)},
scale only axis,
axis on top,
xmin=0.5,
xmax=64.5,
tick align=outside,
y dir=reverse,
ymin=0.5,
ymax=64.5,
axis line style={draw=none},
ticks=none
]
\addplot [forget plot] graphics [xmin=0.5, xmax=64.5, ymin=0.5, ymax=64.5] {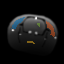};
\end{axis}

\begin{axis}[%
width=3.702in,
height=3.702in,
at={(0in,0in)},
scale only axis,
xmin=0,
xmax=1,
ymin=0,
ymax=1,
axis line style={draw=none},
ticks=none,
axis x line*=bottom,
axis y line*=left
]
\draw[-{Latex}, color=mycolor1, line width=2.0pt] (axis cs:0.80,0.20) -- (axis cs:0.80,0.40);
\draw[-{Latex}, color=mycolor2, line width=2.0pt] (axis cs:0.25,0.15) -- (axis cs:0.40,0.30);
\draw[-{Latex}, color=mycolor3, line width=2.0pt] (axis cs:0.75,0.85) -- (axis cs:0.65,0.65);
\draw[-{Latex}, color=mycolor4, line width=2.0pt] (axis cs:0.20,0.82) -- (axis cs:0.24,0.70);
\node[below, align=center, font=\color{mycolor1}\Large]
at (rel axis cs:0.80,0.17) {\#2: right \\ temporal muscle};
\node[below, align=center, font=\color{mycolor2}\Large]
at (rel axis cs:0.20,0.13) {\#3: tongue};
\node[below, align=center, font=\color{mycolor3}\Large]
at (rel axis cs:0.80,0.93) {\#4: brain tumor};
\node[below, align=center, font=\color{mycolor4}\Large]
at (rel axis cs:0.20,0.97) {\#1: left \\ temporal muscle};

\end{axis}
\end{tikzpicture}%
}
    \caption{Simulated phantom (a~single frame of the ground truth sequence) with indicated ROIs used for evaluation in Sec.\,\ref{ssec:perfusion}.}
    \label{fig:ROI}
\end{figure}
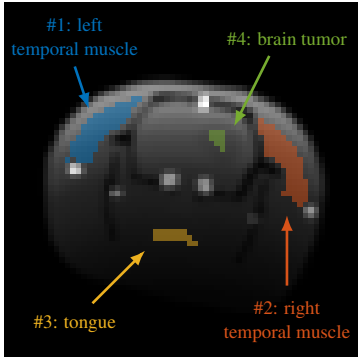

For the purpose of creating the training and testing dataset,
we used an in-house DCE perfusion simulator.
It can produce a~high fidelity synthetic phantom of a rat brain with glioma (see Fig.\,\ref{fig:ROI})
and create undersampled datasets for reconstruction and perfusion analysis testing.
The DCE curves in the phantom were generated using the ATH pharmacokinetic model approximated by the DCATH model 
with a fixed standard deviation of the mean capillary transit times
\cite{Koh2001:DCATH}.
The synthetic echoes were generated using the radial golden angle strategy \cite{Feng2022:Golden.angle.radial.MRI} with TR of 7.5\,ms, TE 1.6\,ms and FA $20^\circ$\!, simulating 4 coils in parallel with smoothed sensitivities from a real scanner.
The image matrix was $64\times64$, and 40,000 projections were simulated.
In order to subsample the dataset,
a~pseudo 3D stack-of-stars acquisition with 16~slices was simulated,
therefore the resulting dataset has 16~times lower temporal resolution compared to plain 2D acquisition. 
%
%

\subsection{Training}

The training dataset consisted of 16 sequences simulated according to Sec.\,\ref{ssec:simulation} with additive Gaussian noise in the k-space with the standard deviation of $\E{1}{-3}$\!.
To define the training loss, we used the ground truth image sequence which was obtained as reconstruction of the simulated data prior to undersampling and adding noise;
the loss was represented by the mean absolute error (MAE) between this ground truth sequence and the reconstructed one.
We trained the network parameters using the Adam optimiser \cite{Kingma2014:Adam} with 200 epochs and learning rate of $\E{2}{-4}$\!.
Even though the aim of the training was to obtain high quality image sequences, we observed that the dynamics of the perfusion curves was better represented when only the first 15\,\% of length of the input sequence was used for training
(i.e.\ 6,000 projections based on the parameters mentioned in Sec.\,\ref{ssec:simulation}).

\begin{table*}[ht]
%
\setlength{\tabcolsep}{9pt}
\centering
\caption{Experiment results.
    Note that the discrepancy between MAE in training and testing is expected since in training,
    the loss was computed only from the first 15 \% of the sequence and it is computed batch-wise,
    while the presented test results were computed on a single (whole) sequence.
}
\label{tab:results}
\vspace{-8pt}
\begin{tabular}{@{}ccccccrcr@{}}
\toprule
\multirow{2}{*}{\begin{tabular}[c]{@{}c@{}}experiment \\ number\end{tabular}} & \multirow{2}{*}{activation} & \multirow{2}{*}{tied} & \multirow{2}{*}{\begin{tabular}[c]{@{}c@{}}layers\\ (iterations)\end{tabular}} & \multirow{2}{*}{\begin{tabular}[c]{@{}c@{}}trained\\ parameters\end{tabular}} & \multicolumn{2}{c}{initialisation} & \multirow{2}{*}{\begin{tabular}[c]{@{}c@{}}training loss\\ (MAE)\end{tabular}} & \multicolumn{1}{c}{\multirow{2}{*}{test MAE}} \\
& & & & & $\lambdaL$ & \multicolumn{1}{c}{$\lambdaS$} & & \multicolumn{1}{c}{} \\ \midrule
\#1 & simple & yes & 100 & 2 & $0.0500$ & $\E{5}{-5}$ & \multicolumn{1}{r}{$\E{2.18}{-4}$} & $\E{7.26}{-5}$ \\
\#2 & simple & no & 100 & 200 & $0.0500$ & $\E{5}{-5}$ & \multicolumn{1}{r}{$\E{1.43}{-4}$} & $\E{5.22}{-5}$ \\
\#3 & soft & yes & 100 & 4 & $0.0234$ & $\E{1.6}{-5}$ & \multicolumn{1}{r}{$\E{1.82}{-4}$} & $\E{7.02}{-5}$ \\
\#4 & soft & no & 100 & 400 & $0.0234$ & $\E{1.6}{-5}$ & \multicolumn{1}{r}{$\E{1.35}{-4}$} & $\E{4.87}{-5}$ \\
\#5 & soft & no & 100 & 400 & $0.0010$ & $\E{1}{-4}$ & \multicolumn{1}{r}{$\E{1.34}{-4}$} & $\E{4.87}{-5}$ \\
\#6 & garrote & no & 100 & 200 & $0.0010$ & $\E{1}{-4}$ & \multicolumn{1}{r}{$\E{1.48}{-4}$} & $\E{5.42}{-5}$ \\
\midrule
\multicolumn{3}{c}{baseline L+S} & 100 & --- & $0.0234$ & $\E{1.6}{-5}$ & --- & $\E{6.68}{-5}$ \\ \bottomrule
\end{tabular}
\end{table*}

\subsection{Testing and evaluation of image sequences}

A straightforward comparison of the results is provided in Table \ref{tab:results}, where the MAE for different setups of the unfolded network is displayed.
A classical CPA for the L+S model (Alg.~\ref{alg:CP})
serves as the baseline  with the regularisation parameters $\lambdaL$ and $\lambdaS$ determined by grid search on the training dataset with the MAE criterion.

Experiment \#1 aims at answering a~question
whether learning the parameters $\lambdaL$ and $\lambdaS$ (shared across all the layers, i.e.\ in the \emph{tied} variant) can be used instead of the grid search.
The learned values were $\lambdaL = 0.0258$ and $\lambdaS = \E{6.8}{-4}$\!, which only seems promising regarding $\lambdaL$.
The reason might be sensitivity of the learning to initialisation and the learning rate.
Furthermore, comparing \#1 with the rest of the experiments, especially with \#2, 
proves that the \emph{tied} approach is too simplistic.
The same observation
stems from the comparison of \#3 with \#4 (the \emph{soft} case).

In experiments \#4 and \#5, the choice of initialisation is examined.
Even though we only test two possibilities, the results suggest that the initial thresholds need not be precisely set to provide a good starting point for learning the \emph{soft} model.
Another observation is that even with a few learnable parameters, the baseline algorithm was outperformed.

Finally, experiment \#6 shows that the choice of the nonnegative garrote is not beneficial,
which suggests that the offset caused by thresholding large input values does not represent a~significant problem in the L+S reconstruction.

\subsection{Evaluation of perfusion maps}
\label{ssec:perfusion}

\begin{figure*}
    \vspace{6pt}
     \centering
     \begin{subfigure}[b]{0.33\textwidth}
         \centering
         \scalebox{0.67}{
%
%
\begin{tikzpicture}[font=\large]

\begin{axis}[%
width=1.906in,
height=1.274in,
at={(0.746in,0.765in)},
scale only axis,
point meta min=3.60510682634365,
point meta max=76.7950939387397,
axis on top,
xmin=0.5,
xmax=4.5,
xtick={1,2,3,4},
xticklabels={{ROI \#1},{ROI \#2},{ROI \#3},{ROI \#4}},
xticklabel style={rotate=45},
y dir=reverse,
ymin=0.5,
ymax=5.5,
ytick={1,2,3,4,5},
yticklabels={{\Fp},{$E$},{\Tc},{\Ktrans},{$PS$}},
axis background/.style={fill=white},
colormap={mymap}{[1pt] rgb(0pt)=(0.2422,0.1504,0.6603); rgb(1pt)=(0.2444,0.1534,0.6728); rgb(2pt)=(0.2464,0.1569,0.6847); rgb(3pt)=(0.2484,0.1607,0.6961); rgb(4pt)=(0.2503,0.1648,0.7071); rgb(5pt)=(0.2522,0.1689,0.7179); rgb(6pt)=(0.254,0.1732,0.7286); rgb(7pt)=(0.2558,0.1773,0.7393); rgb(8pt)=(0.2576,0.1814,0.7501); rgb(9pt)=(0.2594,0.1854,0.761); rgb(11pt)=(0.2628,0.1932,0.7828); rgb(12pt)=(0.2645,0.1972,0.7937); rgb(13pt)=(0.2661,0.2011,0.8043); rgb(14pt)=(0.2676,0.2052,0.8148); rgb(15pt)=(0.2691,0.2094,0.8249); rgb(16pt)=(0.2704,0.2138,0.8346); rgb(17pt)=(0.2717,0.2184,0.8439); rgb(18pt)=(0.2729,0.2231,0.8528); rgb(19pt)=(0.274,0.228,0.8612); rgb(20pt)=(0.2749,0.233,0.8692); rgb(21pt)=(0.2758,0.2382,0.8767); rgb(22pt)=(0.2766,0.2435,0.884); rgb(23pt)=(0.2774,0.2489,0.8908); rgb(24pt)=(0.2781,0.2543,0.8973); rgb(25pt)=(0.2788,0.2598,0.9035); rgb(26pt)=(0.2794,0.2653,0.9094); rgb(27pt)=(0.2798,0.2708,0.915); rgb(28pt)=(0.2802,0.2764,0.9204); rgb(29pt)=(0.2806,0.2819,0.9255); rgb(30pt)=(0.2809,0.2875,0.9305); rgb(31pt)=(0.2811,0.293,0.9352); rgb(32pt)=(0.2813,0.2985,0.9397); rgb(33pt)=(0.2814,0.304,0.9441); rgb(34pt)=(0.2814,0.3095,0.9483); rgb(35pt)=(0.2813,0.315,0.9524); rgb(36pt)=(0.2811,0.3204,0.9563); rgb(37pt)=(0.2809,0.3259,0.96); rgb(38pt)=(0.2807,0.3313,0.9636); rgb(39pt)=(0.2803,0.3367,0.967); rgb(40pt)=(0.2798,0.3421,0.9702); rgb(41pt)=(0.2791,0.3475,0.9733); rgb(42pt)=(0.2784,0.3529,0.9763); rgb(43pt)=(0.2776,0.3583,0.9791); rgb(44pt)=(0.2766,0.3638,0.9817); rgb(45pt)=(0.2754,0.3693,0.984); rgb(46pt)=(0.2741,0.3748,0.9862); rgb(47pt)=(0.2726,0.3804,0.9881); rgb(48pt)=(0.271,0.386,0.9898); rgb(49pt)=(0.2691,0.3916,0.9912); rgb(50pt)=(0.267,0.3973,0.9924); rgb(51pt)=(0.2647,0.403,0.9935); rgb(52pt)=(0.2621,0.4088,0.9946); rgb(53pt)=(0.2591,0.4145,0.9955); rgb(54pt)=(0.2556,0.4203,0.9965); rgb(55pt)=(0.2517,0.4261,0.9974); rgb(56pt)=(0.2473,0.4319,0.9983); rgb(57pt)=(0.2424,0.4378,0.9991); rgb(58pt)=(0.2369,0.4437,0.9996); rgb(59pt)=(0.2311,0.4497,0.9995); rgb(60pt)=(0.225,0.4559,0.9985); rgb(61pt)=(0.2189,0.462,0.9968); rgb(62pt)=(0.2128,0.4682,0.9948); rgb(63pt)=(0.2066,0.4743,0.9926); rgb(64pt)=(0.2006,0.4803,0.9906); rgb(65pt)=(0.195,0.4861,0.9887); rgb(66pt)=(0.1903,0.4919,0.9867); rgb(67pt)=(0.1869,0.4975,0.9844); rgb(68pt)=(0.1847,0.503,0.9819); rgb(69pt)=(0.1831,0.5084,0.9793); rgb(70pt)=(0.1818,0.5138,0.9766); rgb(71pt)=(0.1806,0.5191,0.9738); rgb(72pt)=(0.1795,0.5244,0.9709); rgb(73pt)=(0.1785,0.5296,0.9677); rgb(74pt)=(0.1778,0.5349,0.9641); rgb(75pt)=(0.1773,0.5401,0.9602); rgb(76pt)=(0.1768,0.5452,0.956); rgb(77pt)=(0.1764,0.5504,0.9516); rgb(78pt)=(0.1755,0.5554,0.9473); rgb(79pt)=(0.174,0.5605,0.9432); rgb(80pt)=(0.1716,0.5655,0.9393); rgb(81pt)=(0.1686,0.5705,0.9357); rgb(82pt)=(0.1649,0.5755,0.9323); rgb(83pt)=(0.161,0.5805,0.9289); rgb(84pt)=(0.1573,0.5854,0.9254); rgb(85pt)=(0.154,0.5902,0.9218); rgb(86pt)=(0.1513,0.595,0.9182); rgb(87pt)=(0.1492,0.5997,0.9147); rgb(88pt)=(0.1475,0.6043,0.9113); rgb(89pt)=(0.1461,0.6089,0.908); rgb(90pt)=(0.1446,0.6135,0.905); rgb(91pt)=(0.1429,0.618,0.9022); rgb(92pt)=(0.1408,0.6226,0.8998); rgb(93pt)=(0.1383,0.6272,0.8975); rgb(94pt)=(0.1354,0.6317,0.8953); rgb(95pt)=(0.1321,0.6363,0.8932); rgb(96pt)=(0.1288,0.6408,0.891); rgb(97pt)=(0.1253,0.6453,0.8887); rgb(98pt)=(0.1219,0.6497,0.8862); rgb(99pt)=(0.1185,0.6541,0.8834); rgb(100pt)=(0.1152,0.6584,0.8804); rgb(101pt)=(0.1119,0.6627,0.877); rgb(102pt)=(0.1085,0.6669,0.8734); rgb(103pt)=(0.1048,0.671,0.8695); rgb(104pt)=(0.1009,0.675,0.8653); rgb(105pt)=(0.0964,0.6789,0.8609); rgb(106pt)=(0.0914,0.6828,0.8562); rgb(107pt)=(0.0855,0.6865,0.8513); rgb(108pt)=(0.0789,0.6902,0.8462); rgb(109pt)=(0.0713,0.6938,0.8409); rgb(110pt)=(0.0628,0.6972,0.8355); rgb(111pt)=(0.0535,0.7006,0.8299); rgb(112pt)=(0.0433,0.7039,0.8242); rgb(113pt)=(0.0328,0.7071,0.8183); rgb(114pt)=(0.0234,0.7103,0.8124); rgb(115pt)=(0.0155,0.7133,0.8064); rgb(116pt)=(0.0091,0.7163,0.8003); rgb(117pt)=(0.0046,0.7192,0.7941); rgb(118pt)=(0.0019,0.722,0.7878); rgb(119pt)=(0.0009,0.7248,0.7815); rgb(120pt)=(0.0018,0.7275,0.7752); rgb(121pt)=(0.0046,0.7301,0.7688); rgb(122pt)=(0.0094,0.7327,0.7623); rgb(123pt)=(0.0162,0.7352,0.7558); rgb(124pt)=(0.0253,0.7376,0.7492); rgb(125pt)=(0.0369,0.74,0.7426); rgb(126pt)=(0.0504,0.7423,0.7359); rgb(127pt)=(0.0638,0.7446,0.7292); rgb(128pt)=(0.077,0.7468,0.7224); rgb(129pt)=(0.0899,0.7489,0.7156); rgb(130pt)=(0.1023,0.751,0.7088); rgb(131pt)=(0.1141,0.7531,0.7019); rgb(132pt)=(0.1252,0.7552,0.695); rgb(133pt)=(0.1354,0.7572,0.6881); rgb(134pt)=(0.1448,0.7593,0.6812); rgb(135pt)=(0.1532,0.7614,0.6741); rgb(136pt)=(0.1609,0.7635,0.6671); rgb(137pt)=(0.1678,0.7656,0.6599); rgb(138pt)=(0.1741,0.7678,0.6527); rgb(139pt)=(0.1799,0.7699,0.6454); rgb(140pt)=(0.1853,0.7721,0.6379); rgb(141pt)=(0.1905,0.7743,0.6303); rgb(142pt)=(0.1954,0.7765,0.6225); rgb(143pt)=(0.2003,0.7787,0.6146); rgb(144pt)=(0.2061,0.7808,0.6065); rgb(145pt)=(0.2118,0.7828,0.5983); rgb(146pt)=(0.2178,0.7849,0.5899); rgb(147pt)=(0.2244,0.7869,0.5813); rgb(148pt)=(0.2318,0.7887,0.5725); rgb(149pt)=(0.2401,0.7905,0.5636); rgb(150pt)=(0.2491,0.7922,0.5546); rgb(151pt)=(0.2589,0.7937,0.5454); rgb(152pt)=(0.2695,0.7951,0.536); rgb(153pt)=(0.2809,0.7964,0.5266); rgb(154pt)=(0.2929,0.7975,0.517); rgb(155pt)=(0.3052,0.7985,0.5074); rgb(156pt)=(0.3176,0.7994,0.4975); rgb(157pt)=(0.3301,0.8002,0.4876); rgb(158pt)=(0.3424,0.8009,0.4774); rgb(159pt)=(0.3548,0.8016,0.4669); rgb(160pt)=(0.3671,0.8021,0.4563); rgb(161pt)=(0.3795,0.8026,0.4454); rgb(162pt)=(0.3921,0.8029,0.4344); rgb(163pt)=(0.405,0.8031,0.4233); rgb(164pt)=(0.4184,0.803,0.4122); rgb(165pt)=(0.4322,0.8028,0.4013); rgb(166pt)=(0.4463,0.8024,0.3904); rgb(167pt)=(0.4608,0.8018,0.3797); rgb(168pt)=(0.4753,0.8011,0.3691); rgb(169pt)=(0.4899,0.8002,0.3586); rgb(170pt)=(0.5044,0.7993,0.348); rgb(171pt)=(0.5187,0.7982,0.3374); rgb(172pt)=(0.5329,0.797,0.3267); rgb(173pt)=(0.547,0.7957,0.3159); rgb(175pt)=(0.5748,0.7929,0.2941); rgb(176pt)=(0.5886,0.7913,0.2833); rgb(177pt)=(0.6024,0.7896,0.2726); rgb(178pt)=(0.6161,0.7878,0.2622); rgb(179pt)=(0.6297,0.7859,0.2521); rgb(180pt)=(0.6433,0.7839,0.2423); rgb(181pt)=(0.6567,0.7818,0.2329); rgb(182pt)=(0.6701,0.7796,0.2239); rgb(183pt)=(0.6833,0.7773,0.2155); rgb(184pt)=(0.6963,0.775,0.2075); rgb(185pt)=(0.7091,0.7727,0.1998); rgb(186pt)=(0.7218,0.7703,0.1924); rgb(187pt)=(0.7344,0.7679,0.1852); rgb(188pt)=(0.7468,0.7654,0.1782); rgb(189pt)=(0.759,0.7629,0.1717); rgb(190pt)=(0.771,0.7604,0.1658); rgb(191pt)=(0.7829,0.7579,0.1608); rgb(192pt)=(0.7945,0.7554,0.157); rgb(193pt)=(0.806,0.7529,0.1546); rgb(194pt)=(0.8172,0.7505,0.1535); rgb(195pt)=(0.8281,0.7481,0.1536); rgb(196pt)=(0.8389,0.7457,0.1546); rgb(197pt)=(0.8495,0.7435,0.1564); rgb(198pt)=(0.86,0.7413,0.1587); rgb(199pt)=(0.8703,0.7392,0.1615); rgb(200pt)=(0.8804,0.7372,0.165); rgb(201pt)=(0.8903,0.7353,0.1695); rgb(202pt)=(0.9,0.7336,0.1749); rgb(203pt)=(0.9093,0.7321,0.1815); rgb(204pt)=(0.9184,0.7308,0.189); rgb(205pt)=(0.9272,0.7298,0.1973); rgb(206pt)=(0.9357,0.729,0.2061); rgb(207pt)=(0.944,0.7285,0.2151); rgb(208pt)=(0.9523,0.7284,0.2237); rgb(209pt)=(0.9606,0.7285,0.2312); rgb(210pt)=(0.9689,0.7292,0.2373); rgb(211pt)=(0.977,0.7304,0.2418); rgb(212pt)=(0.9842,0.733,0.2446); rgb(213pt)=(0.99,0.7365,0.2429); rgb(214pt)=(0.9946,0.7407,0.2394); rgb(215pt)=(0.9966,0.7458,0.2351); rgb(216pt)=(0.9971,0.7513,0.2309); rgb(217pt)=(0.9972,0.7569,0.2267); rgb(218pt)=(0.9971,0.7626,0.2224); rgb(219pt)=(0.9969,0.7683,0.2181); rgb(220pt)=(0.9966,0.774,0.2138); rgb(221pt)=(0.9962,0.7798,0.2095); rgb(222pt)=(0.9957,0.7856,0.2053); rgb(223pt)=(0.9949,0.7915,0.2012); rgb(224pt)=(0.9938,0.7974,0.1974); rgb(225pt)=(0.9923,0.8034,0.1939); rgb(226pt)=(0.9906,0.8095,0.1906); rgb(227pt)=(0.9885,0.8156,0.1875); rgb(228pt)=(0.9861,0.8218,0.1846); rgb(229pt)=(0.9835,0.828,0.1817); rgb(230pt)=(0.9807,0.8342,0.1787); rgb(231pt)=(0.9778,0.8404,0.1757); rgb(232pt)=(0.9748,0.8467,0.1726); rgb(233pt)=(0.972,0.8529,0.1695); rgb(234pt)=(0.9694,0.8591,0.1665); rgb(235pt)=(0.9671,0.8654,0.1636); rgb(236pt)=(0.9651,0.8716,0.1608); rgb(237pt)=(0.9634,0.8778,0.1582); rgb(238pt)=(0.9619,0.884,0.1557); rgb(239pt)=(0.9608,0.8902,0.1532); rgb(240pt)=(0.9601,0.8963,0.1507); rgb(241pt)=(0.9596,0.9023,0.148); rgb(242pt)=(0.9595,0.9084,0.145); rgb(243pt)=(0.9597,0.9143,0.1418); rgb(244pt)=(0.9601,0.9203,0.1382); rgb(245pt)=(0.9608,0.9262,0.1344); rgb(246pt)=(0.9618,0.932,0.1304); rgb(247pt)=(0.9629,0.9379,0.1261); rgb(248pt)=(0.9642,0.9437,0.1216); rgb(249pt)=(0.9657,0.9494,0.1168); rgb(250pt)=(0.9674,0.9552,0.1116); rgb(251pt)=(0.9692,0.9609,0.1061); rgb(252pt)=(0.9711,0.9667,0.1001); rgb(253pt)=(0.973,0.9724,0.0938); rgb(254pt)=(0.9749,0.9782,0.0872); rgb(255pt)=(0.9769,0.9839,0.0805)},
colorbar,
colorbar style={ytick={10, 30, 50, 70}, yticklabel={\pgfmathparse{\tick}\pgfmathprintnumber{\pgfmathresult} \%}}
]
\addplot [forget plot] graphics [xmin=0.5, xmax=4.5, ymin=0.5, ymax=5.5] {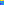};
\end{axis}

\end{tikzpicture}%
}
         \caption{baseline relative errors}
         \label{fig:perfusion:gtc_1}
     \end{subfigure}
     \hfill
     \begin{subfigure}[b]{0.33\textwidth}
         \centering
         \scalebox{0.67}{
%
%
\begin{tikzpicture}[font=\large]

\begin{axis}[%
width=1.906in,
height=1.274in,
at={(0.746in,0.765in)},
scale only axis,
point meta min=3.60510682634365,
point meta max=76.7950939387397,
axis on top,
xmin=0.5,
xmax=4.5,
xtick={1,2,3,4},
xticklabels={{ROI \#1},{ROI \#2},{ROI \#3},{ROI \#4}},
xticklabel style={rotate=45},
y dir=reverse,
ymin=0.5,
ymax=5.5,
ytick={1,2,3,4,5},
yticklabels={{\Fp},{$E$},{\Tc},{\Ktrans},{$PS$}},
axis background/.style={fill=white},
colormap={mymap}{[1pt] rgb(0pt)=(0.2422,0.1504,0.6603); rgb(1pt)=(0.2444,0.1534,0.6728); rgb(2pt)=(0.2464,0.1569,0.6847); rgb(3pt)=(0.2484,0.1607,0.6961); rgb(4pt)=(0.2503,0.1648,0.7071); rgb(5pt)=(0.2522,0.1689,0.7179); rgb(6pt)=(0.254,0.1732,0.7286); rgb(7pt)=(0.2558,0.1773,0.7393); rgb(8pt)=(0.2576,0.1814,0.7501); rgb(9pt)=(0.2594,0.1854,0.761); rgb(11pt)=(0.2628,0.1932,0.7828); rgb(12pt)=(0.2645,0.1972,0.7937); rgb(13pt)=(0.2661,0.2011,0.8043); rgb(14pt)=(0.2676,0.2052,0.8148); rgb(15pt)=(0.2691,0.2094,0.8249); rgb(16pt)=(0.2704,0.2138,0.8346); rgb(17pt)=(0.2717,0.2184,0.8439); rgb(18pt)=(0.2729,0.2231,0.8528); rgb(19pt)=(0.274,0.228,0.8612); rgb(20pt)=(0.2749,0.233,0.8692); rgb(21pt)=(0.2758,0.2382,0.8767); rgb(22pt)=(0.2766,0.2435,0.884); rgb(23pt)=(0.2774,0.2489,0.8908); rgb(24pt)=(0.2781,0.2543,0.8973); rgb(25pt)=(0.2788,0.2598,0.9035); rgb(26pt)=(0.2794,0.2653,0.9094); rgb(27pt)=(0.2798,0.2708,0.915); rgb(28pt)=(0.2802,0.2764,0.9204); rgb(29pt)=(0.2806,0.2819,0.9255); rgb(30pt)=(0.2809,0.2875,0.9305); rgb(31pt)=(0.2811,0.293,0.9352); rgb(32pt)=(0.2813,0.2985,0.9397); rgb(33pt)=(0.2814,0.304,0.9441); rgb(34pt)=(0.2814,0.3095,0.9483); rgb(35pt)=(0.2813,0.315,0.9524); rgb(36pt)=(0.2811,0.3204,0.9563); rgb(37pt)=(0.2809,0.3259,0.96); rgb(38pt)=(0.2807,0.3313,0.9636); rgb(39pt)=(0.2803,0.3367,0.967); rgb(40pt)=(0.2798,0.3421,0.9702); rgb(41pt)=(0.2791,0.3475,0.9733); rgb(42pt)=(0.2784,0.3529,0.9763); rgb(43pt)=(0.2776,0.3583,0.9791); rgb(44pt)=(0.2766,0.3638,0.9817); rgb(45pt)=(0.2754,0.3693,0.984); rgb(46pt)=(0.2741,0.3748,0.9862); rgb(47pt)=(0.2726,0.3804,0.9881); rgb(48pt)=(0.271,0.386,0.9898); rgb(49pt)=(0.2691,0.3916,0.9912); rgb(50pt)=(0.267,0.3973,0.9924); rgb(51pt)=(0.2647,0.403,0.9935); rgb(52pt)=(0.2621,0.4088,0.9946); rgb(53pt)=(0.2591,0.4145,0.9955); rgb(54pt)=(0.2556,0.4203,0.9965); rgb(55pt)=(0.2517,0.4261,0.9974); rgb(56pt)=(0.2473,0.4319,0.9983); rgb(57pt)=(0.2424,0.4378,0.9991); rgb(58pt)=(0.2369,0.4437,0.9996); rgb(59pt)=(0.2311,0.4497,0.9995); rgb(60pt)=(0.225,0.4559,0.9985); rgb(61pt)=(0.2189,0.462,0.9968); rgb(62pt)=(0.2128,0.4682,0.9948); rgb(63pt)=(0.2066,0.4743,0.9926); rgb(64pt)=(0.2006,0.4803,0.9906); rgb(65pt)=(0.195,0.4861,0.9887); rgb(66pt)=(0.1903,0.4919,0.9867); rgb(67pt)=(0.1869,0.4975,0.9844); rgb(68pt)=(0.1847,0.503,0.9819); rgb(69pt)=(0.1831,0.5084,0.9793); rgb(70pt)=(0.1818,0.5138,0.9766); rgb(71pt)=(0.1806,0.5191,0.9738); rgb(72pt)=(0.1795,0.5244,0.9709); rgb(73pt)=(0.1785,0.5296,0.9677); rgb(74pt)=(0.1778,0.5349,0.9641); rgb(75pt)=(0.1773,0.5401,0.9602); rgb(76pt)=(0.1768,0.5452,0.956); rgb(77pt)=(0.1764,0.5504,0.9516); rgb(78pt)=(0.1755,0.5554,0.9473); rgb(79pt)=(0.174,0.5605,0.9432); rgb(80pt)=(0.1716,0.5655,0.9393); rgb(81pt)=(0.1686,0.5705,0.9357); rgb(82pt)=(0.1649,0.5755,0.9323); rgb(83pt)=(0.161,0.5805,0.9289); rgb(84pt)=(0.1573,0.5854,0.9254); rgb(85pt)=(0.154,0.5902,0.9218); rgb(86pt)=(0.1513,0.595,0.9182); rgb(87pt)=(0.1492,0.5997,0.9147); rgb(88pt)=(0.1475,0.6043,0.9113); rgb(89pt)=(0.1461,0.6089,0.908); rgb(90pt)=(0.1446,0.6135,0.905); rgb(91pt)=(0.1429,0.618,0.9022); rgb(92pt)=(0.1408,0.6226,0.8998); rgb(93pt)=(0.1383,0.6272,0.8975); rgb(94pt)=(0.1354,0.6317,0.8953); rgb(95pt)=(0.1321,0.6363,0.8932); rgb(96pt)=(0.1288,0.6408,0.891); rgb(97pt)=(0.1253,0.6453,0.8887); rgb(98pt)=(0.1219,0.6497,0.8862); rgb(99pt)=(0.1185,0.6541,0.8834); rgb(100pt)=(0.1152,0.6584,0.8804); rgb(101pt)=(0.1119,0.6627,0.877); rgb(102pt)=(0.1085,0.6669,0.8734); rgb(103pt)=(0.1048,0.671,0.8695); rgb(104pt)=(0.1009,0.675,0.8653); rgb(105pt)=(0.0964,0.6789,0.8609); rgb(106pt)=(0.0914,0.6828,0.8562); rgb(107pt)=(0.0855,0.6865,0.8513); rgb(108pt)=(0.0789,0.6902,0.8462); rgb(109pt)=(0.0713,0.6938,0.8409); rgb(110pt)=(0.0628,0.6972,0.8355); rgb(111pt)=(0.0535,0.7006,0.8299); rgb(112pt)=(0.0433,0.7039,0.8242); rgb(113pt)=(0.0328,0.7071,0.8183); rgb(114pt)=(0.0234,0.7103,0.8124); rgb(115pt)=(0.0155,0.7133,0.8064); rgb(116pt)=(0.0091,0.7163,0.8003); rgb(117pt)=(0.0046,0.7192,0.7941); rgb(118pt)=(0.0019,0.722,0.7878); rgb(119pt)=(0.0009,0.7248,0.7815); rgb(120pt)=(0.0018,0.7275,0.7752); rgb(121pt)=(0.0046,0.7301,0.7688); rgb(122pt)=(0.0094,0.7327,0.7623); rgb(123pt)=(0.0162,0.7352,0.7558); rgb(124pt)=(0.0253,0.7376,0.7492); rgb(125pt)=(0.0369,0.74,0.7426); rgb(126pt)=(0.0504,0.7423,0.7359); rgb(127pt)=(0.0638,0.7446,0.7292); rgb(128pt)=(0.077,0.7468,0.7224); rgb(129pt)=(0.0899,0.7489,0.7156); rgb(130pt)=(0.1023,0.751,0.7088); rgb(131pt)=(0.1141,0.7531,0.7019); rgb(132pt)=(0.1252,0.7552,0.695); rgb(133pt)=(0.1354,0.7572,0.6881); rgb(134pt)=(0.1448,0.7593,0.6812); rgb(135pt)=(0.1532,0.7614,0.6741); rgb(136pt)=(0.1609,0.7635,0.6671); rgb(137pt)=(0.1678,0.7656,0.6599); rgb(138pt)=(0.1741,0.7678,0.6527); rgb(139pt)=(0.1799,0.7699,0.6454); rgb(140pt)=(0.1853,0.7721,0.6379); rgb(141pt)=(0.1905,0.7743,0.6303); rgb(142pt)=(0.1954,0.7765,0.6225); rgb(143pt)=(0.2003,0.7787,0.6146); rgb(144pt)=(0.2061,0.7808,0.6065); rgb(145pt)=(0.2118,0.7828,0.5983); rgb(146pt)=(0.2178,0.7849,0.5899); rgb(147pt)=(0.2244,0.7869,0.5813); rgb(148pt)=(0.2318,0.7887,0.5725); rgb(149pt)=(0.2401,0.7905,0.5636); rgb(150pt)=(0.2491,0.7922,0.5546); rgb(151pt)=(0.2589,0.7937,0.5454); rgb(152pt)=(0.2695,0.7951,0.536); rgb(153pt)=(0.2809,0.7964,0.5266); rgb(154pt)=(0.2929,0.7975,0.517); rgb(155pt)=(0.3052,0.7985,0.5074); rgb(156pt)=(0.3176,0.7994,0.4975); rgb(157pt)=(0.3301,0.8002,0.4876); rgb(158pt)=(0.3424,0.8009,0.4774); rgb(159pt)=(0.3548,0.8016,0.4669); rgb(160pt)=(0.3671,0.8021,0.4563); rgb(161pt)=(0.3795,0.8026,0.4454); rgb(162pt)=(0.3921,0.8029,0.4344); rgb(163pt)=(0.405,0.8031,0.4233); rgb(164pt)=(0.4184,0.803,0.4122); rgb(165pt)=(0.4322,0.8028,0.4013); rgb(166pt)=(0.4463,0.8024,0.3904); rgb(167pt)=(0.4608,0.8018,0.3797); rgb(168pt)=(0.4753,0.8011,0.3691); rgb(169pt)=(0.4899,0.8002,0.3586); rgb(170pt)=(0.5044,0.7993,0.348); rgb(171pt)=(0.5187,0.7982,0.3374); rgb(172pt)=(0.5329,0.797,0.3267); rgb(173pt)=(0.547,0.7957,0.3159); rgb(175pt)=(0.5748,0.7929,0.2941); rgb(176pt)=(0.5886,0.7913,0.2833); rgb(177pt)=(0.6024,0.7896,0.2726); rgb(178pt)=(0.6161,0.7878,0.2622); rgb(179pt)=(0.6297,0.7859,0.2521); rgb(180pt)=(0.6433,0.7839,0.2423); rgb(181pt)=(0.6567,0.7818,0.2329); rgb(182pt)=(0.6701,0.7796,0.2239); rgb(183pt)=(0.6833,0.7773,0.2155); rgb(184pt)=(0.6963,0.775,0.2075); rgb(185pt)=(0.7091,0.7727,0.1998); rgb(186pt)=(0.7218,0.7703,0.1924); rgb(187pt)=(0.7344,0.7679,0.1852); rgb(188pt)=(0.7468,0.7654,0.1782); rgb(189pt)=(0.759,0.7629,0.1717); rgb(190pt)=(0.771,0.7604,0.1658); rgb(191pt)=(0.7829,0.7579,0.1608); rgb(192pt)=(0.7945,0.7554,0.157); rgb(193pt)=(0.806,0.7529,0.1546); rgb(194pt)=(0.8172,0.7505,0.1535); rgb(195pt)=(0.8281,0.7481,0.1536); rgb(196pt)=(0.8389,0.7457,0.1546); rgb(197pt)=(0.8495,0.7435,0.1564); rgb(198pt)=(0.86,0.7413,0.1587); rgb(199pt)=(0.8703,0.7392,0.1615); rgb(200pt)=(0.8804,0.7372,0.165); rgb(201pt)=(0.8903,0.7353,0.1695); rgb(202pt)=(0.9,0.7336,0.1749); rgb(203pt)=(0.9093,0.7321,0.1815); rgb(204pt)=(0.9184,0.7308,0.189); rgb(205pt)=(0.9272,0.7298,0.1973); rgb(206pt)=(0.9357,0.729,0.2061); rgb(207pt)=(0.944,0.7285,0.2151); rgb(208pt)=(0.9523,0.7284,0.2237); rgb(209pt)=(0.9606,0.7285,0.2312); rgb(210pt)=(0.9689,0.7292,0.2373); rgb(211pt)=(0.977,0.7304,0.2418); rgb(212pt)=(0.9842,0.733,0.2446); rgb(213pt)=(0.99,0.7365,0.2429); rgb(214pt)=(0.9946,0.7407,0.2394); rgb(215pt)=(0.9966,0.7458,0.2351); rgb(216pt)=(0.9971,0.7513,0.2309); rgb(217pt)=(0.9972,0.7569,0.2267); rgb(218pt)=(0.9971,0.7626,0.2224); rgb(219pt)=(0.9969,0.7683,0.2181); rgb(220pt)=(0.9966,0.774,0.2138); rgb(221pt)=(0.9962,0.7798,0.2095); rgb(222pt)=(0.9957,0.7856,0.2053); rgb(223pt)=(0.9949,0.7915,0.2012); rgb(224pt)=(0.9938,0.7974,0.1974); rgb(225pt)=(0.9923,0.8034,0.1939); rgb(226pt)=(0.9906,0.8095,0.1906); rgb(227pt)=(0.9885,0.8156,0.1875); rgb(228pt)=(0.9861,0.8218,0.1846); rgb(229pt)=(0.9835,0.828,0.1817); rgb(230pt)=(0.9807,0.8342,0.1787); rgb(231pt)=(0.9778,0.8404,0.1757); rgb(232pt)=(0.9748,0.8467,0.1726); rgb(233pt)=(0.972,0.8529,0.1695); rgb(234pt)=(0.9694,0.8591,0.1665); rgb(235pt)=(0.9671,0.8654,0.1636); rgb(236pt)=(0.9651,0.8716,0.1608); rgb(237pt)=(0.9634,0.8778,0.1582); rgb(238pt)=(0.9619,0.884,0.1557); rgb(239pt)=(0.9608,0.8902,0.1532); rgb(240pt)=(0.9601,0.8963,0.1507); rgb(241pt)=(0.9596,0.9023,0.148); rgb(242pt)=(0.9595,0.9084,0.145); rgb(243pt)=(0.9597,0.9143,0.1418); rgb(244pt)=(0.9601,0.9203,0.1382); rgb(245pt)=(0.9608,0.9262,0.1344); rgb(246pt)=(0.9618,0.932,0.1304); rgb(247pt)=(0.9629,0.9379,0.1261); rgb(248pt)=(0.9642,0.9437,0.1216); rgb(249pt)=(0.9657,0.9494,0.1168); rgb(250pt)=(0.9674,0.9552,0.1116); rgb(251pt)=(0.9692,0.9609,0.1061); rgb(252pt)=(0.9711,0.9667,0.1001); rgb(253pt)=(0.973,0.9724,0.0938); rgb(254pt)=(0.9749,0.9782,0.0872); rgb(255pt)=(0.9769,0.9839,0.0805)},
colorbar,
colorbar style={ytick={10, 30, 50, 70}, yticklabel={\pgfmathparse{\tick}\pgfmathprintnumber{\pgfmathresult} \%}}
]
\addplot [forget plot] graphics [xmin=0.5, xmax=4.5, ymin=0.5, ymax=5.5] {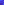};
\end{axis}

\end{tikzpicture}%
}
         \caption{trained model relative errors}
         \label{fig:perfusion:gtc_2}
     \end{subfigure}
     \hfill
     \begin{subfigure}[b]{0.33\textwidth}
         \centering
         \scalebox{0.67}{\input{figures/barvicky_1699273006_gtc_3.tex}}
         \caption{trained vs.\ baseline (less is better)}
         \label{fig:perfusion:gtc_3}
     \end{subfigure}
    %
    \caption{Comparison of the perfusion map quality, the baseline L+S algorithm (\subref{fig:perfusion:gtc_1}) versus the trained model from experiment \#4 (\subref{fig:perfusion:gtc_2}).
    The relative error in a single pixel of a perfusion map is computed as the absolute difference of the examined value and the reference value, divided by the absolute value of the reference.
    In each ROI and per each perfusion parameter considered, the relative errors are then averaged (see also Fig.\,\ref{fig:ROI}).
    Note that the graph in (\subref{fig:perfusion:gtc_3}) shows the difference of percentages, which is why the percentage point is used as a unit of the colorbar.
    For example, the average relative error in terms of \Fp{} in ROI \#1 is around 60\,\% for the baseline and around 14\,\% for the trained model, which results in the value of around $-46$\,\%pt.\ in
    graph (\subref{fig:perfusion:gtc_3}), i.e.\ a large improvement over the baseline.
    Note that a symmetric colormap is used on purpose
    to enable comparison with Fig.\,\ref{fig:perfusion:sim_3}, even though all the values in \subref{fig:perfusion:gtc_3} are negative (blue).
    }
    \label{fig:perfusion}
    \vspace{-3pt}
\end{figure*}

\begin{figure}[h]
    \centering
    \scalebox{0.67}{\input{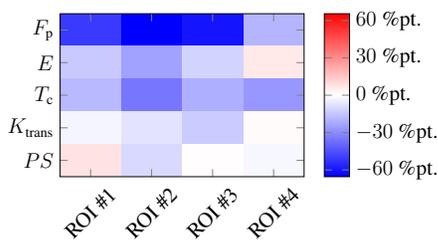}}
    \caption{Difference of the errors plotted in the same way as in Fig.\,\ref{fig:perfusion:gtc_3},
    but for the option of comparison to the perfusion maps used to simulate the DCE-MRI data.
    Only the differences are shown for the sake of brevity,
    however, the actual error values are in similar range as in graphs \ref{fig:perfusion:gtc_1} and \ref{fig:perfusion:gtc_2}.}
    \label{fig:perfusion:sim_3}
    \vspace{-12pt}
\end{figure}

To evaluate the perfusion map quality, relevant ROIs for diagnostics were chosen.
That is three muscles (typically used for arterial input function scale calibration) and the tumor itself (see Fig.\,\ref{fig:ROI}).
We mainly focused on the \Ktrans{} parameter for its diagnostic relevancy and on \Fp{} as one of the more advanced parameters to estimate.
Due to the fact that simulated data was used,
there are two options how to evaluate
the quality of the perfusion maps.
The first choice is to employ the raw maps used during simulation.
However, comparison with these maps may be distorted by the process of automated perfusion analysis \cite{perflab}.
Furthermore, there are inevitable physics-related limitations in the acquisition (e.g.\ the coil sensitivity) which are not reflected in these ground truth maps.
For this reason, the second option may be preferable, where the
perfusion analysis is performed using the \emph{reconstructed} ground truth image sequence to produce the reference maps.

A comparison using the latter option is presented in Fig.\,\ref{fig:perfusion}.
The results indicate a consistent improvement of the perfusion analysis.
In particular, significant improvement is observed in the \Fp{} parameter which cannot be well estimated from the baseline reconstruction (see first line of the graph \ref{fig:perfusion:gtc_1}), but the error in the estimation of \Fp{} using the learned reconstruction is comparable to the estimation errors in other parameters.

An abridged evaluation using the simulated maps is shown in Fig.\,\ref{fig:perfusion:sim_3}.
In contrast to the evaluation using the maps estimated from the ground truth image sequence, an increase in the error is observed here in a few cases.
However, looking and the results globally, reconstruction using the trained model is favoured also in this evaluation.

\section{Conclusion}

The paper introduced a simple unfolded scheme for the reconstruction of DCE-MRI data, based on the L+S model.
The results indicate that even simple parametrisation of the model allows to improve quality of the reconstructed sequence in terms of absolute error, compared to the classical L+S baseline.
Remarkably, the experiments also demonstrated that this
kind of
objective quality also coincides with the quality improvement in the perfusion maps.
This offers a~conclusion that even with a simple network (based on a well-designed model) and simple loss function,
the perfusion analysis can be enhanced.

A further improvement may be reached by designing a~temporal-weighted loss function for the training,
which would preserve differentiability while allowing to capture
the parts of the sequence most important for the perfusion analysis.
Another future direction is the focus on more shallow network with more parameters per layer.
Since interpretability and theoretical guarantees are substantial, divergence from the proximal L+S model is not desired, which suggests for example the use of (suitably constrained) adaptive piecewise-linear activation \cite{Agostinelli2014:APL} instead of the thresholding operators.


\vspace{-3pt}
\section{Compliance with ethical standards}
\label{sec:ethics}
\vspace{-3pt}

This is a numerical simulation study for which no ethical approval was required.

\vspace{-3pt}
\section{Acknowledgements}
\label{sec:acknowledgments}
\vspace{-3pt}

The work was supported by the Czech Science Foundation
(GA\v{C}R) Project No.\,22-10953S and project no.\,LM2023050 of the MEYS CR.


\vspace{-3pt}

\begin{thebibliography}{10}

\bibitem{Chambolle2004:TV.algorithm}
Antonin Chambolle,
\newblock ``An algorithm for total variation minimization and applications,''
\newblock {\em Journal of Mathematical Imaging and Vision}, vol. 20, no. 1--2, 2004.

\vspace{-4pt}

\bibitem{Schloegl_ICTGV}
Matthias Schloegl, Martin Holler, Andreas Schwarzl, Kristian Bredies, and Rudolf Stollberger,
\newblock ``Infimal convolution of total generalized variation functionals for dynamic {MRI},''
\newblock {\em Magnetic Resonance in Medicine}, vol. 78, no. 1, 2017.

\vspace{-4pt}

\bibitem{Dankova2016}
Marie Daňková and Pavel Rajmic,
\newblock ``Low-rank model for dynamic {MRI}: joint solving and debiasing,''
\newblock in {\em 
33rd Annual Scientific Meeting: Magnetic resonance materials in physics biology and medicine}. 2016, Berlin: Springer.

\vspace{-4pt}

\bibitem{MangovaTSP2017}
Marie Mangová, Pavel Rajmic, and Radovan Jiřík,
\newblock ``Dynamic magnetic resonance imaging using compressed sensing with multi-scale low rank penalty,''
\newblock in {\em 40th International Conference on Telecommunications and Signal Processing}, 2017.

\vspace{-4pt}

\bibitem{Candes:2011Robust.princ.comp.anal}
Emmanuel~J. Cand\`{e}s, Xiaodong Li, Yi~Ma, and John Wright,
\newblock ``Robust principal component analysis?,''
\newblock {\em J. ACM}, vol. 58, no. 3, 2011.

\vspace{-4pt}

\bibitem{OtazoCandesSodickson:MRIperfusion}
Ricardo Otazo, Emmanuel~J. Cand\`{e}s, and Daniel~K. Sodickson,
\newblock ``Low-rank plus sparse matrix decomposition for accelerated dynamic {MRI} with separation of background and dynamic components,''
\newblock {\em Magnetic Resonance in Medicine}, vol. 73, no. 3, 2015.

\vspace{-4pt}

\bibitem{DankovaRajmicJirik2015:LVA}
Marie Daňková, Pavel Rajmic, and Radovan Jiřík,
\newblock ``Acceleration of perfusion {MRI} using locally low-rank plus sparse model,''
\newblock in {\em Latent Variable Analysis and Signal Separation}, Liberec, 2015, Springer.

\vspace{-4pt}

\bibitem{Schlemper2017:Deep.cascade.of.CNN.for.MRI}
Jo~Schlemper, Jose Caballero, Joseph~V. Hajnal, Anthony Price, and Daniel Rueckert,
\newblock ``A deep cascade of convolutional neural networks for {MR} image reconstruction,''
\newblock in {\em Lecture Notes in Computer Science}. Springer International Publishing, 2017.

\vspace{-4pt}

\bibitem{Knoll2020:Deep.Learning.Parallel.MRI.Survey}
Florian Knoll, Kerstin Hammernik, Chi Zhang, Steen Moeller, Thomas Pock et al.
\newblock ``Deep-learning methods for parallel magnetic resonance imaging reconstruction: A survey of the current approaches, trends, and issues,''
\newblock {\em {IEEE} Signal Processing Magazine}, vol. 37, no. 1, 2020.

\vspace{-4pt}

\bibitem{Liang2020:Deep.MRI}
Dong Liang, Jing Cheng, Ziwen Ke, and Leslie Ying,
\newblock ``Deep magnetic resonance image reconstruction: Inverse problems meet neural networks,''
\newblock {\em {IEEE} Signal Processing Magazine}, vol. 37, no. 1, 2020.

\vspace{-4pt}

\bibitem{Hammernik2017:Learning.Variational.Network.MRI}
Kerstin Hammernik, Teresa Klatzer, Erich Kobler, Michael~P. Recht, Daniel~K. Sodickson et al.
\newblock ``Learning a variational network for reconstruction of accelerated {MRI} data,''
\newblock {\em Magnetic Resonance in Medicine}, vol. 79, no. 6, 2017.

\vspace{-4pt}

\bibitem{Huang2021:Deep.L.plus.S.network}
Wenqi Huang, Ziwen Ke, Zhuo-Xu Cui, Jing Cheng, Zhilang Qiu et al.
\newblock ``Deep low-rank plus sparse network for dynamic {MR} imaging,''
\newblock {\em Medical Image Analysis}, vol. 73, 2021.

\vspace{-4pt}

\bibitem{Mokry2022:MRI.EEICT}
Ondřej Mokrý and Jiří Vitouš,
\newblock ``Unfolded low-rank + sparse reconstruction for {MRI},''
\newblock in {\em Proceedings II of the 28th Conference {STUDENT EEICT} 2022 Selected papers}, Brno, 2022, vol.~1.

\vspace{-4pt}

\bibitem{Recht2010:nuclear.norm}
B.~Recht, M.~Fazel, and P.~Parrilo,
\newblock ``Guaranteed minimum-rank solutions of linear matrix equations via nuclear norm minimization,''
\newblock {\em SIAM Review}, vol. 52, no. 3, 2010.

\vspace{-4pt}

\bibitem{DonohoElad2003:Optimally}
David~L. Donoho and Michael Elad,
\newblock ``Optimally sparse representation in general (nonorthogonal) dictionaries via $\ell_1$ minimization,''
\newblock {\em Proceedings of The National Academy of Sciences}, vol. 100, no. 5, 2003.

\vspace{-4pt}

\bibitem{ChambollePock2011:First-Order.Primal-Dual.Algorithm}
Antonin Chambolle and Thomas Pock,
\newblock ``A first-order primal-dual algorithm for convex problems with applications to imaging,''
\newblock {\em Journal of Mathematical Imaging and Vision}, vol. 40, no. 1, 2011.

\vspace{-4pt}

\bibitem{Cai2010:SVT}
Jian-Feng Cai, Emmanuel~J. Cand{\`{e}}s, and Zuowei Shen,
\newblock ``A singular value thresholding algorithm for matrix completion,''
\newblock {\em {SIAM} Journal on Optimization}, vol. 20, no. 4, 2010.

\vspace{-4pt}

\bibitem{Liu2019:Deep.Proximal.Unrolling}
Risheng Liu, Shichao Cheng, Long Ma, Xin Fan, and Zhongxuan Luo,
\newblock ``Deep proximal unrolling: Algorithmic framework, convergence analysis and applications,''
\newblock {\em {IEEE} Transactions on Image Processing}, vol. 28, no. 10, 2019.

\vspace{-4pt}

\bibitem{MongaLiEldar2021:Algorithm.unrolling}
Vishal Monga, Yuelong Li, and Yonina~C. Eldar,
\newblock ``Algorithm unrolling: Interpretable, efficient deep learning for signal and image processing,''
\newblock {\em IEEE Signal Processing Magazine}, vol. 38, no. 2, 2021.

\vspace{-4pt}

\bibitem{Gribonval2020:Characterization.of.prox}
R\'emi Gribonval and Mila Nikolova,
\newblock ``A characterization of proximity operators,''
\newblock {\em Journal of Mathematical Imaging and Vision}, vol. 62, no. 6--7, 2020.

\vspace{-4pt}

\bibitem{Breiman1995:Nonnegative.garrote}
Leo Breiman,
\newblock ``Better subset regression using the nonnegative garrote,''
\newblock {\em Technometrics}, vol. 37, no. 4, 1995.

\vspace{-4pt}

\bibitem{Koh2001:DCATH}
T.~S. Koh, V.~Zeman, J.~Darko, T.-Y. Lee, M.~F. Milosevic et al.
\newblock ``The inclusion of capillary distribution in the adiabatic tissue homogeneity model of blood flow,''
\newblock {\em Physics in Medicine and Biology}, vol. 46, no. 5, 2001.

\vspace{-4pt}

\bibitem{Feng2022:Golden.angle.radial.MRI}
Li~Feng,
\newblock ``Golden-angle radial {MRI}: Basics, advances, and applications,''
\newblock {\em Journal of Magnetic Resonance Imaging}, vol. 56, no. 1, 2022.

\vspace{-4pt}

\bibitem{Kingma2014:Adam}
Diederik~P. Kingma and Jimmy Ba,
\newblock ``Adam: A method for stochastic optimization,'' 2014.
\url{https://arxiv.org/abs/1412.6980}

\vspace{-4pt}

\bibitem{perflab}
``Perfusion lab,'' online. \url{http://perflab.cerit-sc.cz/}. Accessed 10 November 2023.

\vspace{-4pt}

\bibitem{Agostinelli2014:APL}
Forest Agostinelli, Matthew Hoffman, Peter Sadowski, and Pierre Baldi,
\newblock ``Learning activation functions to improve deep neural networks,'' 2014.
\url{https://arxiv.org/abs/1412.6830}

\end{thebibliography}

\newcommand{\noopsort}[1]{} \newcommand{\printfirst}[2]{#1} \newcommand{\singleletter}[1]{#1} \newcommand{\switchargs}[2]{#2#1}

\end{document}